\documentclass[preprint,showpacs,preprintnumbers,amsmath,amssymb]{revtex4}

\usepackage{graphicx}
\usepackage{dcolumn}
\usepackage{bm}

\begin{document}

\title{Large-N scaling behavior of the ground-state energy, fidelity, and the order parameter  in the
Dicke Model}
\author{Tao Liu$^{1,2}$, Yu-Yu Zhang$^{2,3}$, Qing-Hu Chen$^{2,3,*}$, and Ke-Lin
Wang $^{1,4}$}
\address{
$^{1}$ Department of Physics, Southwest University of  Science and
Technology, Mianyang 621010, P.  R.  China\\
$^2$ Center for Statistical and Theoretical Condensed Matter
Physics, Zhejiang Normal University, Jinhua 321004, P. R. China \\
$^{3}$ Department of Physics, Zhejiang University, Hangzhou 310027,
P. R. China \\
$^{4}$Department of Modern Physics, University of  Science and
Technology of China,  Hefei 230026, P.  R.  China
 }

\date{\today}

\begin{abstract}
Within the numerically exact solution to the Dicke model proposed
previously, we study the quantum criticality in terms of the
ground-state (GS) energy, fidelity, and the order parameter. The
finite size scaling analysis for the average fidelity susceptibility
(FS) and second derivative of GS energy are performed. The
correlation length exponent is obtained to be $\nu=2/3$, which is
the same as that in Lipkin-Meshkov-Glick model obtained previously,
suggesting the same universality. It is observed that average FS and
second derivative of GS energy show similar critical behavior,
demonstrating the intrinsic relation in the Dicke model. The scaling
behavior for the order parameter and the  singular part of the GS
energy at the critical point are also analyzed and the obtained
exponents are consistent with the previous scaling hypothesis in 1/N
expansion scheme.
\end{abstract}


\maketitle

\draft

\section{introduction}

The Dicke model\cite{dicke} describes the interaction of N two-level
atoms (qubits) with a single bosonic mode, which is a fundamental
model in quantum optics. It exhibits a "superrandiant" quantum phase
transition (QPT)\cite {Sachdev} in the thermodynamic limit. In
recent years, the Dicke model has attracted considerable
attentions\cite
{Emary,Lambert1,Buzek,Lambert2,liberti,vidal,reslen,plastina,chen1}.
On the one hand, the quantum entanglement\cite{oster} and Berry
phase \cite{Carollo} have been used to characterize the QPTs. On the
other hand, the Dicke model is closely related to many recent
interesting fields in quantum optics and condensed matter physics,
such as the superradiant behavior by an ensemble of quantum dots
\cite{Scheibner} and Bose-Einstein condensates \cite {Schneble}, and
coupled arrays of optical cavities\cite{Hartmann}.

The  scaling exponents obtained at the critical points are of
significance to distinguish the universality class of the QPTs among
various models.  By a  modified Holstein-Primakoff approach, Vidal
and  Dusuel has predicted theoretically the nontrivial scaling
exponent for several quantities in the Dicke model\cite{vidal}. To
our knowledge, the finite-size studies in the Dicke model were
previously limited to numerical diagonalization in Bosonic Fock
state \cite{Emary,Lambert1,Lambert2} in small size system $N\leq
35$, the adiabatic approximation \cite{liberti}. Recently, by using
extended bosonic coherent states,  a numerically exact technique to
solve the Dicke model for large system size was developed by the
present authors\cite{chen1}. Therefore, numerical calculations of
scaling exponents for some key quantities based on convincing
treatment for large system size are  clearly desirable for
confirmation.

Recently, a concept in quantum information theory, i.e. the fidelity
has been extensively used to identify the QPTs in various many-body
systems from the perspective of the ground-state (GS) wave
functions\cite
{Quan,Zanardi,Cozzini,You,Gu,Shu,zhou,Wang,sun,Venuti,Kowk} (For
more details, please refer to a review article \cite{review}). In a
mathematical sense, the fidelity is the overlap between two ground
states where the transition parameters deviate slightly. However,
the fidelity depends on a arbitrary small amount of the transition
parameters, which in turn yields an artificial factor. Zanardi et al
\cite{Cozzini} introduced the Riemannian metric tensor and You et al
\cite{You} proposed the fidelity susceptibility (FS) to avoid this
problem independently. The leading term of the fidelity are focused
to account for the singularity of QPTs in both methods. It is thus
implied that the fidelity have no singular behavior in the
Kosterliz-Thouless phase transition, which can distinguish different
transition types\cite{Gu,Shu,Venuti}. In addition, the intrinsic
relation between the GS fidelity and the derivative of GS energy has
been studied and it is observed that they play a equivalent role in
identifying the QPTs\cite {Shu}. Since no \emph{a priori} knowledge
of the order parameter is needed, it may be a great advantage to
study the scaling behavior of FS to characterize the universality in
quantum critical phenomena\cite{Gu}.

In this paper, we extended our previous numerically exact technique
to calculate the GS energy, FS, and the order parameter in the Dicke
model for finite N. The scaling behavior of FS, the second
derivative of GS energy, and the order parameter of the QPT will be
analyzed. We will also study the singular part of the GS energy,
which is a further correction to the regular part \cite{vidal}.

\section{Model Hamiltonian}

Without the rotating-wave approximation, the Hamiltonian of $N$ identical
atoms interacting with a single bosonic mode reads
\begin{equation}
H=\omega a^{+}a+\omega _0J_z+\frac{2\lambda }{\sqrt{N}}(a^{+}+a)J_x,
\label{hamiltonian}
\end{equation}
where $a^{+}$ and $a$ are the bosonic annihilation and creation operators, $%
\omega _0$ and $\omega $ are the transition frequency of the qubit and the
frequency of the single bosonic mode, $\lambda $ is the coupling constant. $%
J_x$ and $J_z$ are the usual angular momentum. There is a conserved parity
operator $\Pi =e^{i\pi (Jz+N/2+a^{+}a)}$, which commutes with the
Hamiltonian (1).

In our previous numerically exact approach \cite{chen1}, the wave
function can be expressed in terms of the basis $\{\left| \varphi
_n\right\rangle _b\bigotimes \left| j,n\right\rangle \}$ where $
\left| j,n\right\rangle $ $\ $ is the Dicke state with $j=N/2$ and
$\left| \varphi _n\right\rangle _b$ is the bosonic state. The latter
is given by
\begin{equation}
\left| \varphi _n\right\rangle _b=\sum_{k=0}^{N_{tr}}c_{n,k}\frac 1{\sqrt{k!}
}(a^{+}+g_n)^ke^{-g_na^{+}-g_n^2/2}\left| 0\right\rangle _a,
\label{wavefunction}
\end{equation}
where $g_n=2\lambda n/\omega \sqrt{N}$, $N_{tr}$ is the truncated
bosonic number in the Fock space of the new operator $A=a+g_n$, the
coefficient $ c_{n,k}$ can be determined through the exact Lanczos
diagonalization\cite {tian}.

The driving Hamiltonian in the Dicke model is
\begin{equation}
H_1=\frac{2}{\sqrt{N}}(a^{+}+a)J_x,
\end{equation}
and the transition parameter is $\lambda$. According to the definition\cite
{You}, the FS is given by
\begin{equation}
S_F(\lambda )=\sum\limits_{n\neq 0}\frac{|\langle \Psi _n|H_1|\Psi _0\rangle
|^2}{[E_n-E_0]^2}
\end{equation}
where $\Psi _n$ is the n-th eigen-states of the Hamiltonian (1). In terms of
Eq. (2), we have
\begin{eqnarray*}
\langle \Psi _n|H_1|\Psi _0\rangle &=&\frac 2{\sqrt{N}}\sum\limits_{m,i,j}(
\frac N2-m)C_{m,i}^{n*}C_{m,j}^0[\sqrt{j+1}\delta _{i,j+1} \\
&&+\sqrt{j}\delta _{i,j-1}-2g_m\delta _{i,j}].
\end{eqnarray*}
After the FS is well defined above, the fidelity is readily obtained through
an artificially introduced small amount of the transition parameter $\delta
\lambda$
\begin{equation}
F(\lambda ,\delta \lambda )=\sqrt{1-\delta \lambda ^2S_F(\lambda )}
\end{equation}
Although the FS might not be a intensive quantity in some
cases\cite{Gu}, for convenience, we will discuss the average FS:
$\chi _F=S_F(\lambda )/N$ throughout the paper. It is interesting to
note that the FS can be obtained through the  wave function without
the knowledge of  the order parameter.

\section{Finite-size scaling analysis}

We have studied the leading finite-size corrections to the GS energy in the
Dicke model\cite{chen1}. For convenience, we have introduced one
dimensionless parameter $D=\omega _0/\omega $. The QPT occurs at the
critical point $\lambda _c=0.5$ in the thermodynamic limit\cite{Emary}. The
scaling exponent of the leading finite-size corrections is obtained to be $%
-1.0\pm 0.02$ for three typical different values $D=0.1,1,10$, consistent
with that by a modified Holstein-Primakoff approach\cite{vidal}. In this
paper, we first study the scaling behavior of the singular part of the GS
energy, which can be obtained by subtracting the leading contribution \cite
{vidal} $c0+c1/N$, with $c_0=-\omega _0/2$ and $c_1=\frac 12[-\omega -\omega
_0+(\omega ^2+\omega _0^2)^{1/2}]$ from the total GS energy. We exhibit the
curve of $e_0-(c_0+ \frac{c_1}N)$ as a function of $N$ at the critical point
for different typical values of $D$ in Fig. 1. One can obviously observe
that the scaling exponent estimated from the slope in all cases are $%
-1.325\pm 0.005$, which is very close to $-4/3$, confirming the previous
theoretical prediction in a modified Holstein-Primakoff approach\cite{vidal}.

\begin{figure}[tbp]
\centering
\includegraphics[width=8cm]{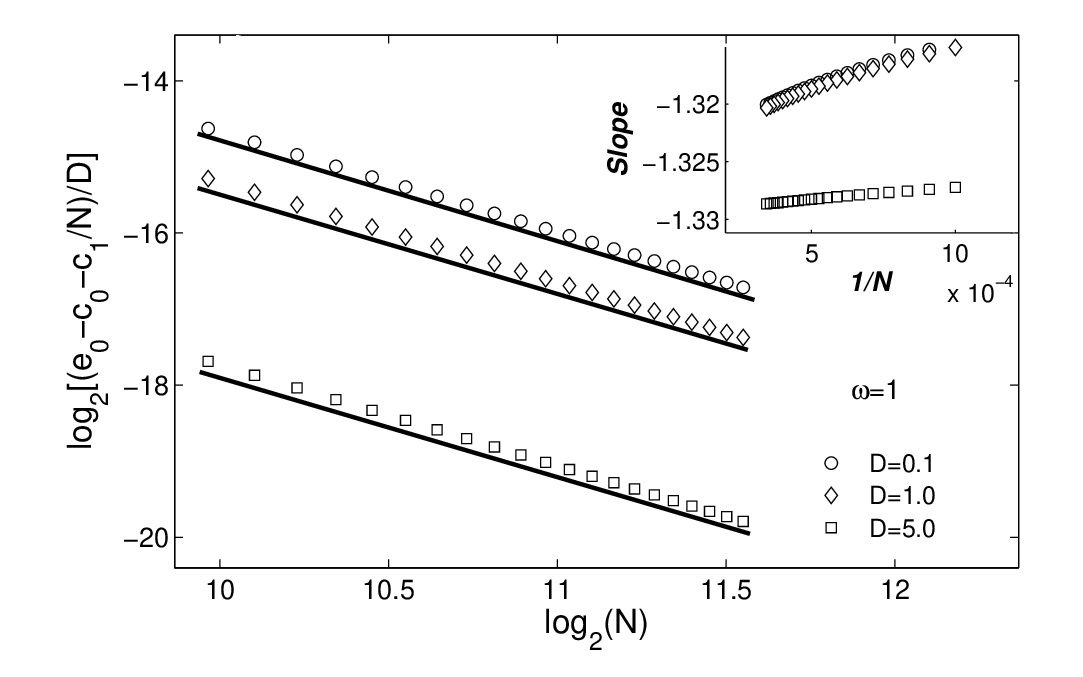}
\caption{Scaling of the singular part of the GS energy $e_0-(c_0+\frac{c_1}
N) $ as a function of $N$ at the critical point for $D=0.1,1$, and $5$. The
inset shows the corresponding slope versus $1/N$. }
\label{fig1}
\end{figure}

Next, we illustrate the scaling behavior of the average FS. The finite-size
scaling ansatz for the average FS to analyze the QPT take the form\cite{Kowk}
\begin{equation}
\frac{\chi _F^{\max }-\chi _F}{\chi _F}=f[N^\nu (\lambda -\lambda _{\max })]
\end{equation}
where $\chi _F^{\max }$ is the value of average FS at the maximum point $%
\lambda _{\max }$, $f$ is the scaling function and $\nu $ is the
correlation length critical exponent. This function should be
universal for large N in the second-order QPTs, which is independent
of the order parameter. As shown in Fig. 2, with $\nu =2/3$, an
excellent collapse in the critical regime is achieved according to
Eq.(6) in the curve for different large size for three typical
values of $D$. Beyond the critical regime, the collapse becomes
poor. As N increases, the critical regime become wider. It is
demonstrated that $\nu $ is a universal constant and does not
depended on the parameter $D$. It is also implied from the collapse
that the correlation length behaves like $\xi \varpropto \left|
\lambda -\lambda_c \right| ^{-2/3}$. It is very interesting to note
that the value of $\nu $ is the same as that in the
Lipkin-Meshkov-Glick (LMG) model obtained analytically \cite{Dusuel}
and numerically\cite{Kowk}, suggesting the same universality in the
Dicke and LMG model.

\begin{figure}[tbp]
\centering
\includegraphics[width=8cm]{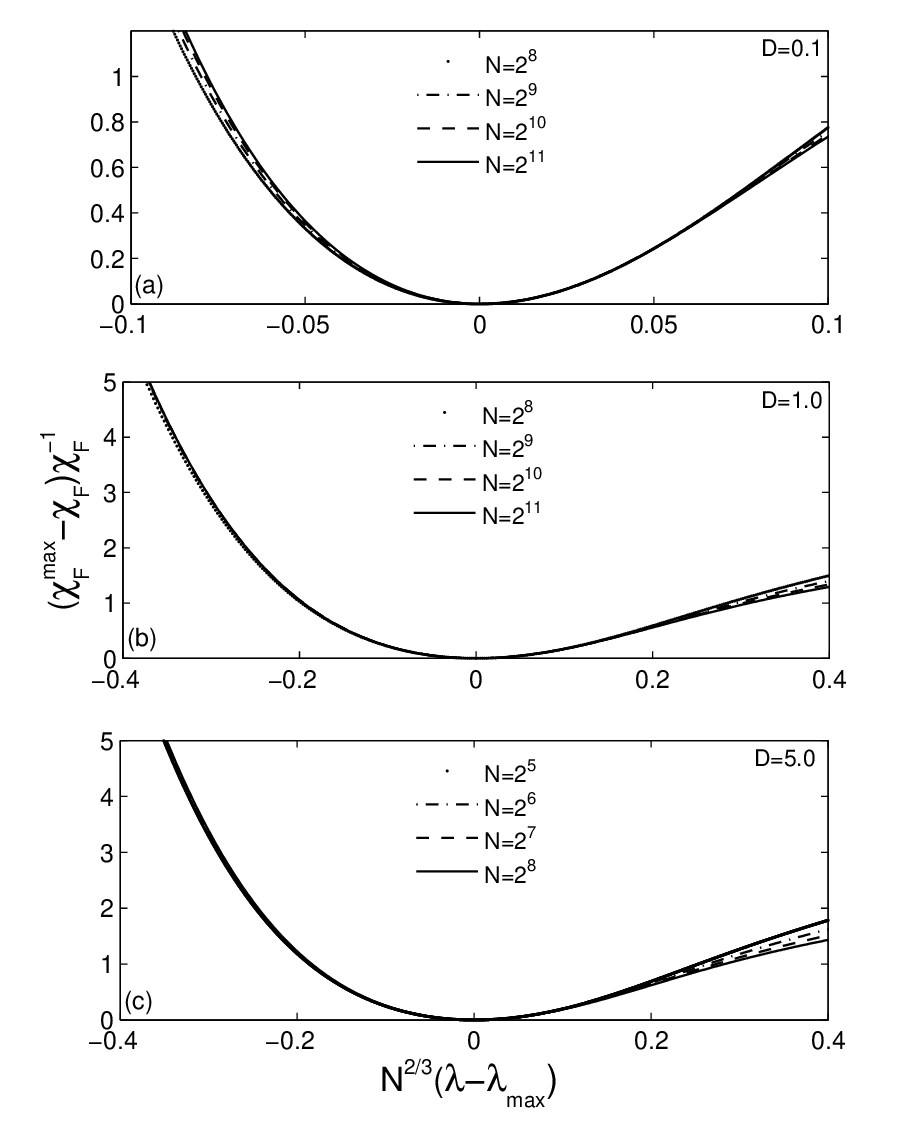}
\caption{Finite-size scaling of the average FS according to Eq. (6) at the
critical point for (a) $D=0.1$, (b) $D=1$, and (c) $D=5$. }
\label{fig2}
\end{figure}

Fig. 3 shows the scaled average FS at the maximum point as a function of $N$
for different values of $D$ in log-log scale. A power law behavior $\chi
_F^{\max }\varpropto N^\mu $ exists in the large $N$. The finite-size
exponents extracted from all curves tend to a converging value $\mu =0.33\pm
0.02$. This exponent is also independent of the value of $D$, and then is a
universal constant. It is also very close to that in the LMG model,
providing another evidence of the same universality class of these two
models.

\begin{figure}[tbp]
\centering
\includegraphics[width=8cm]{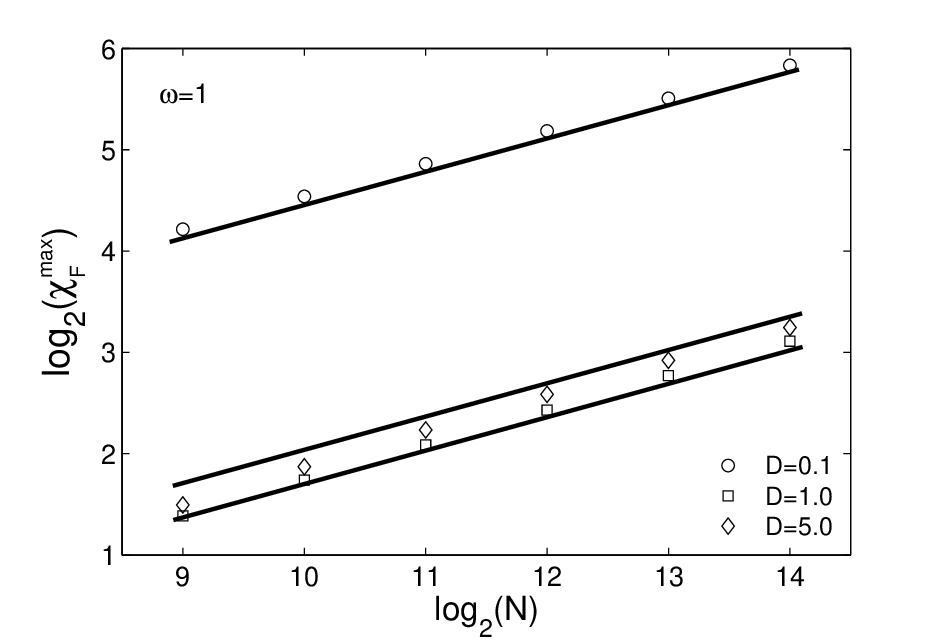}
\caption{Scaling of the maximum of the average FS as a function of $N$ at
the critical point for $D=0.1, 1$, and $5$. The inset shows the
corresponding slope versus $1/N$. }
\label{fig3}
\end{figure}

To exhibit the overall properties of the average FS in the whole coupling
regime, we calculate the average FS as a function of $\lambda $ for
different size. From Fig. 4, we can see that $S_F=N\chi _F $ is a intensive
quantity when $\lambda >\lambda _c$ , while $\chi _F $ is a intensive
quantity when $\lambda <\lambda _c$. It is pointed out in Ref. \cite{Gu,Kowk}
that the intensive average FS in the thermodynamic limit scales generally
like
\begin{equation}
\chi _F\varpropto \frac 1{\left| \lambda -\lambda _c\right| ^\alpha }
\end{equation}
in the vicinity of the critical point. Within the similar analysis, we can
get the exponent$\ \alpha =1/2$ when $\lambda >\lambda _c$ and $\alpha =2$
when $\lambda <\lambda _c\ $ through the relation $\alpha =\mu /\nu ,$ which
is readily derived from the above scaling ansatz. Also it is interesting to
note that the average FS becomes divergent as the system size increases,
demonstrating a Landau-type transitions in the Dicke model.

\begin{figure}[tbp]
\centering
\includegraphics[width=8cm]{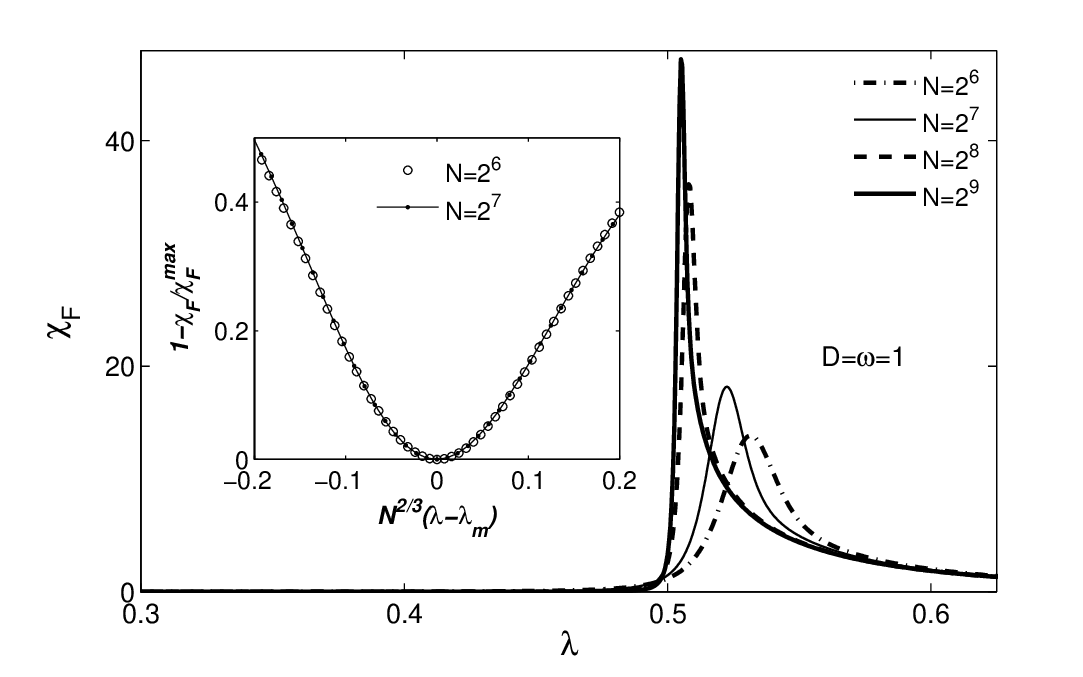}
\caption{The average FS as a function of $\lambda$ at $D=1$ for
different system sizes. The inset shows the collapse of $1-\chi
_F/\chi _F^{\max }$ as a function of $
N^{2/3}(\lambda-\lambda_{max}) $.} \label{fig4}
\end{figure}
To illustrate the intrinsic relation between the ground-state
fidelity and the ground state energy $e_0$ in the Dicke model, we
calculate the second derivative of the GS energy for different
lattice size at $D=1$, which are shown in Fig. 5. Interestingly,
similar to the average FS, the second derivative of the GS energy
also shows divergent behavior around the critical point with
increasing system size. This divergent behavior is also similar to
that observed in 1D transverse field Ising model\cite{Shu}. The only
difference is that the maximum point $ \lambda_{max}$ approaches the
critical point with the increasing system size in the Dicke model
more fast than in the 1D transverse field Ising model, due to the
smaller value of the correlation length exponent $\nu$ in the former
model.

Because the scaling $\lambda -\lambda _{\max }\varpropto N^{-2/3}\ $ holds,
we try to plot the dimensionless quantities $\ 1-\vartheta /\vartheta ^{\max
}$, where $\vartheta $ is either the average FS or the second derivative of
the GS energy for different size as a function of $N^{2/3}(\lambda -\lambda
_{\max })\;$in the insets of Figs. 4 and 5. It is interesting to observe
that the second derivative of the GS energy exhibits the same scaling
behavior around the critical point as the average FS. It follows that these
two quantities play the same roles in the characterizing the QPTs in the
Dicke model. In the 1D transverse field Ising model, the value of the
critical exponent $\nu =1$, which is equal to that in the classical
two-dimensional Ising model, the scaling $\lambda -\lambda _{\max
}\varpropto N^{-1}\ $ should be satisfied, which was just used in the Figs.
1 and 2 in Ref. \cite{Shu}.

\begin{figure}[tbp]
\centering
\includegraphics[width=8cm]{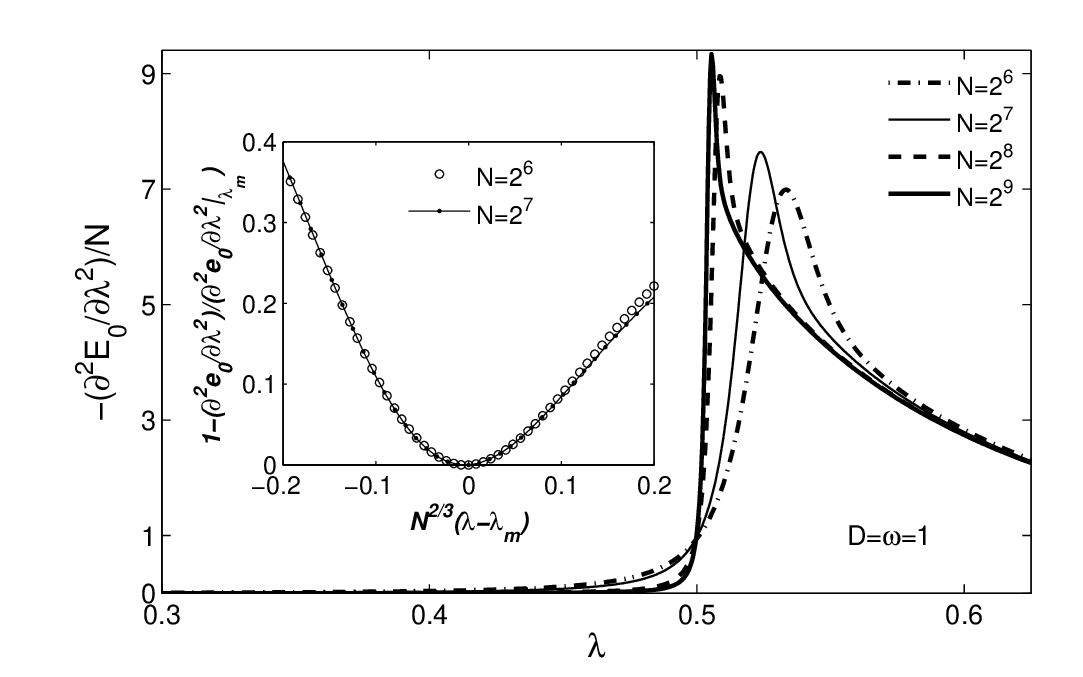}
\caption{The second derivative of the GS energy as a function of
$\lambda $ at $D=1$ for different system sizes. The inset shows the
collapse of $ 1-\frac{\partial ^2\vartheta }{\partial \lambda
^2}/\frac{\partial ^2\vartheta }{\partial \lambda ^2}|_{\lambda _m}
$ as a function of $  N^{2/3}(\lambda-\lambda_{max}) $.}
 \label{fig5}
\end{figure}

We then perform the finite size scaling analysis on the order
parameter of the QPT, i.e. the  expectation value of  the photon
number per atom in the ground-state $ \langle a^{\dagger}a
\rangle/N$. In the thermodynamic limit, this quantity changes from
zero to finite value smoothly  when crossing the critical point. In
Fig. 6, we present this quantity as a function of $N$ for different
values of $D$ in log-log scale. Derivative of these curves are
plotted  in the inset. The exponent of the order parameter is
estimated to be ${-0.66\pm 0.02}$, which is consistent well with the
predicted value $ 2/3$ \cite{vidal} by using the diagonalizing a
expanded Hamiltonian at order $1/N$ based on Holstein-Primakoff
representation.

\begin{figure}[tbp]
\centering
\includegraphics[width=8cm]{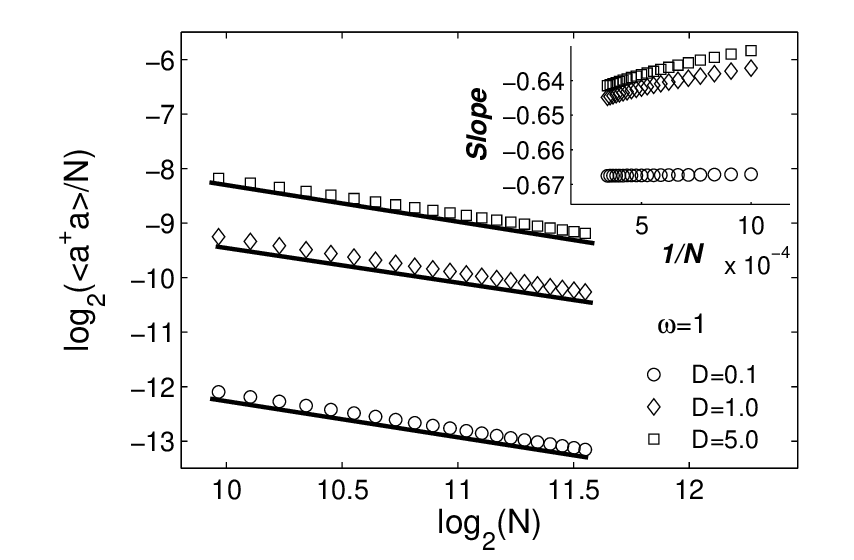}
\caption{Scaling of the order parameter $ \langle a^{\dagger}a
\rangle/N$ as a function of $N$ at the critical point for $D=0.1,
1$, and $5$. The  inset shows the slope versus $1/N$.} \label{fig6}
\end{figure}

Finally,  to describe the new  numerically exact  approach with
extended coherent states in more intuitive detail and see how it
connects to results in the thermodynamic limit, we calculate the
finite but large N wave functions, as done in Ref. \cite{Emary}. We
calculate the ground-state wave function $\Psi(x,y)$  in the x-y
representation for $N=1000$ and $D=1$ at various  $\lambda$ which
cover $\lambda_{c} (N=1000)$. $\lambda_{c}(N=1000)$ falls in the
range of $[0.5070,0.5071]$, according to the position of the peak
for FS. After tedious calculations, the numerical results for the
wave function are shown in Fig. 7, where the displacement is not
removed. One can observe that the wave packet becomes stretched in a
direction with an angle around $\pi/4$ as $\lambda$ increases. This
stretching increases up to $\lambda_{c} (N)$, where the wave
function is not divergent due to the finite-size effect. Exceeding
$\lambda_c$, the wave function splits into two peaked one.  It is
also a finite-size effect and  not shown in the thermodynamic
limit[cf. Fig. 6 in Ref. \cite{Emary}]. With the further increase of
$\lambda$, the two lobes start mixing, a single-peaked wave function
is then formed  like in the thermodynamic limit\cite{Emary}. It
follows that the essential feature in the thermodynamic limit
already appears for large $N$, say $N=1000$. Note that in Fig. 12 of
Ref. \cite{Emary}, two lobes still separate even at $\lambda=0.7$
for $j=5$ (i.e. $N=10$), different from the present observation for
$N=1000$.

\begin{figure}[tbp]
\centering
\includegraphics[width=16cm]{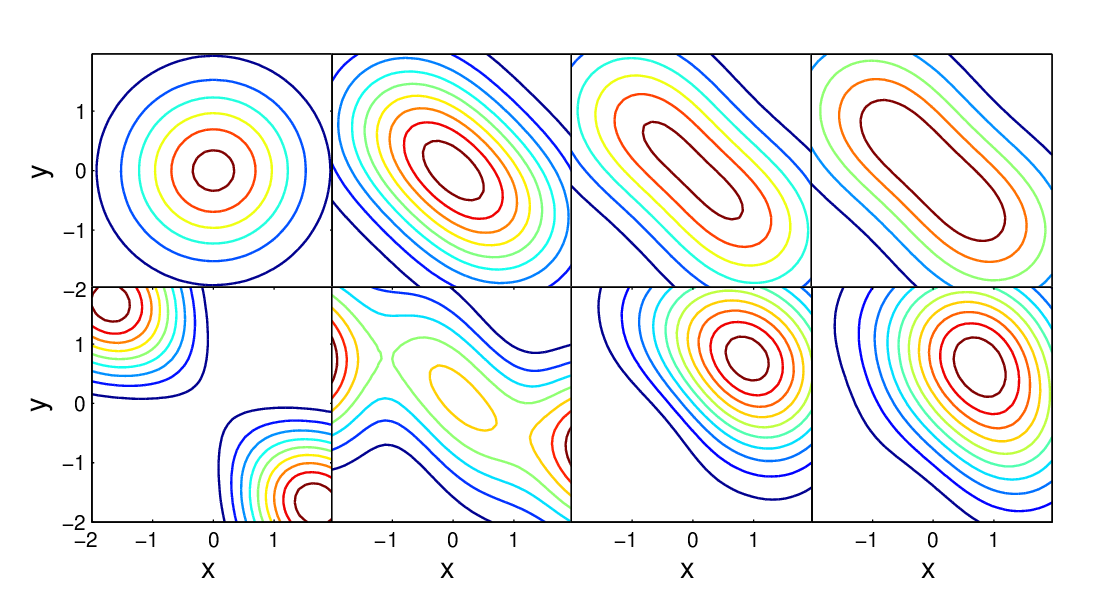}
\caption{The ground-state wave function $\Psi(x,y)$  in the  x-y
representation for $N=1000$ and $D=1$ at $\lambda=0, 0.5, 0.507,
0.5071$ (from left to right in the upper row), $0.525, 0.55, 0.555,
0.7$ (the low row). Note that $0.5070<\lambda_{c} (N=1000)<
0.5071$.}
 \label{fig6}
\end{figure}

\section{Conclusion}

In summary, based on our previously proposed numerically exact
technique in the finite-size the Dicke model, we study the quantum
criticality in terms of the GS energy, fidelity, and the order
parameter. By subtracting the regular part of the GS energy, we
perform the analysis for the scaling behavior of for the singular
part of the GS energy at the critical point. The obtained exponent
is  very close to $4/3$, which agree well with the previous
theoretical prediction in 1/N expansion scheme based on the
Holstein-Primakoff transformation.  Then, we perform the finite-size
scaling analysis for the average FS. Several scaling exponents are
obtained: $\nu =2/3$, $\mu =4/3$, and $\alpha =1/2$ when $\lambda
>\lambda _c$ and $\alpha =2$ when $\lambda <\lambda _c\ $ . All this
exponents are the same as those in Lipkin-Meshkov-Glick model
obtained both analytically  and numerically, suggesting the same
universality of these two models. We further study the scaling
behavior of the average FS and the second derivative of GS energy,
and observed that these two quantities play the same roles in the
QPTs in the Dicke model. Brief comparisons with the 1D transverse
field Ising model are also carried out. We  also perform the
analysis of the finite size scaling effect of the order parameter,
and observe that the order parameter vanishes as $N^{-2/3}$ at the
critical point, consistent with the previous theoretical prediction.
Finally, we calculate the finite but large N wave functions. It is
observed that the essential feature in the thermodynamic limit has
already shown up in large system size ($N=1000$).

\section{Ackownledgements}

The authors acknowledge useful discussion with  S. J. Gu and  J.
Vidal. This work was supported by National Natural Science
Foundation of China, PCSIRT (Grant No. IRT0754) in University in
China, National Basic Research Program of China (Grant No.
2009CB929104), Zhejiang Provincial Natural Science Foundation under
Grant No. Z7080203, and Program for Innovative Research  Team in
Zhejiang Normal University.

$^{*}$ Corresponding author. Email:qhchen@zju.edu.cn


\begin{references}
\bibitem{dicke} R. H. Dicke, Phys. Rev. \textbf{ 93}, 99(1954).
\bibitem{Sachdev} S. Sachdev, Quantum Phase Transitions (Cambridge University Press,
Cambridge, England, 2000).

\bibitem{Emary} C. Emary and T. Brandes, Phys. Rev. E \textbf{67}, 066203(2003); Phys. Rev.
Lett. \textbf{ 90}, 044101(2003).
\bibitem{Lambert1} N. Lambert, C. Emary,  and T. Brandes, Phys. Rev. A. \textbf{ 71},
053804(2005).
\bibitem{Lambert2} N. Lambert, C. Emary,  and T. Brandes, Phys. Rev. Lett. \textbf{ 92},
073602(2004).
\bibitem{liberti} G. Liberti,  F.  Plastina, and F. Piperno,  Phys. Rev. A \textbf{74}, 022324
(2006).
\bibitem{Buzek} V. Buzek, M. Orszag, and M. Rosko,  Phys. Rev. Lett. \textbf{ 94},
163601(2005).
\bibitem{vidal}  J. Vidal and S. Dusuel, Europhys. Lett. \textbf{74},  817(2006).
\bibitem{reslen}  J. Reslen, L. Quiroga, and N. F. Johnson,  Europhys. Lett. \textbf{69},  8(2005).
\bibitem{plastina}  F.  Plastina, G. Liberti, and A. Carollo,
Europhys. Lett. \textbf{ 76},  182(2006).

\bibitem{chen1} Q. H. Chen, Y. Y. Zhang, T.  Liu, and K. L. Wang,
Phys. Rev. A \textbf{78}, 051801(R) (2008).


\bibitem{oster} A. Osterloh, L. Amico, G. Falci, and R. Fazio, Nature
(London) \textbf{416}, 608(2002); T. J. Osborne and M. A. Nielsen,
Phys. Rev. A. \textbf{ 66}, 032110(2002)).
\bibitem{Carollo} A. C. M. Carollo and J. K. Pachos, Phys. Rev. Lett. \textbf{ 95},
157203(2005); S.- L. Zhu, Phys. Rev. Lett. \textbf{ 96},
077206(2006).

\bibitem{Scheibner}  M. Scheibner et al.,  Nature Phys. \textbf{3}, 106(2007).

\bibitem{Schneble} D. Schneble et al.,   Science \textbf{300}, 475
(2003).

\bibitem{Hartmann} M. J. Hartmann et al., Nature Phys. \textbf{2}, 849(2006); A. D. Greentree
et al., ibid. \textbf{2}, 856(2006).

\bibitem{Quan}  H. T. Quan, Z. Song, X. F. Liu, P. Zanardi, and C. P. Sun,
Phys. Rev. Lett. \textbf{96}, 140604 (2006).

\bibitem{Zanardi}  P. Zanardi, and N. Paunkovi\'{c}, Phys. Rev. E. \textbf{74%
}, 0331123 (2006).

\bibitem{Cozzini} P. Zanardi, P.  Giorda, and  M. Cozzini, Phys. Rev. Lett.
\textbf{99}, 100603(2007).

\bibitem{You}  W. L. You, Y. W. Li, and S. J. Gu, Phys. Rev. E. \textbf{76}, 022101 (2007).

\bibitem{Gu} S. J. Gu,  H. M. Kwok, W. Q. Ning, and  H. Q.  Lin,  Phys. Rev. B. \textbf{77}, 245109(2008).

\bibitem{Shu}  S. Chen, L. Wang, Y. J. Hao, and Y. P. Wang, Phys. Rev. A.
\textbf{77}, 032111 (2008).

\bibitem{zhou}  H. Q. Zhou, R. Orus, G. Vidal, Phys. Rev. Lett. \textbf{100}, 080601
(2008); H. Q. Zhou, J. P. Barjaktarevic, arXiv:cond-mat/0701608; H.
Q. Zhou, J. H. Zhao, B. Li, arXiv:0704.2940.

\bibitem{Wang} J. Ma, L. Xu, H.  N. Xiong, and X. G. Wang, Phys. Rev. E. \textbf{78}, 051126
(2008).

\bibitem{sun}  K. W.  Sun, Y. Y.  Zhang, and   Q. H. Chen,  Phys. Rev. B. \textbf{79},  104429
(2009).

\bibitem{Venuti} L. Campos Venuti, M. Cozzini, P. Buonsante, F. Massel, N. Bray-Ali, P.
Zanardi, Phys. Rev. B \textbf{78}, 115410 (2008)

\bibitem{Kowk}  H. M. Kwok, W. Q. Ning, S. J. Gu, and H. Q.  Lin, Phys. Rev. E. \textbf{78}, 032103 (2008).

\bibitem{review} S. J. Gu,  arxiv:quant-ph/0811.3127

\bibitem{tian}G. S. Tian, L. H. Tang, and Q. H. Chen, Europhys. Lett. \textbf{50}, 361(2000);
   Phys. Rev. B63, 054511(2001).


\bibitem{Dusuel} S. Dusuel  and  J. Vidal£¬  Phys. Rev. Lett. \textbf{93}, 237204 (2004);
Phys. Rev. B \textbf{71}, 224420 (2005).



\end{references}
\end{document}